\documentclass{article}

\usepackage{graphicx}
\usepackage{PRIMEarxiv}
\usepackage{natbib}

\def\bSig\mathbf{\Sigma}

\newtheorem{lemma}{Lemma}
\usepackage{color}
\usepackage{amsmath,amssymb}
\usepackage{amsfonts}
\usepackage{booktabs,multirow}  
\usepackage{graphicx,lscape}
\usepackage{mathtools}
\usepackage{bm}
\usepackage{threeparttable}
\usepackage{url}
\usepackage{soul}
\usepackage{subcaption}
\usepackage{enumitem}
\usepackage[justification=raggedright,singlelinecheck=false]{caption}
\numberwithin{equation}{section}
\usepackage{optidef}
\usepackage{ulem} 
\usepackage{graphics}

\usepackage[figuresright]{rotating}

\title{Blinded sample size re-estimation accounting for uncertainty in mid-trial estimation}

\author{Hirotada Maeda \\
Department of Biomedical Statistics, Graduate School of Medicine, The University of Osaka and\\
Department of Data Science, National Cerebral and Cardiovascular Center,\\
Yamadaoka 2-2, Suita City, Osaka 565-0871, Japan\\
\texttt{maeda\_hirotada@biostat.med.osaka-u.ac.jp} \\
	   \And 
	   Satoshi Hattori\\
       Department of Biomedical Statistics, Graduate School of Medicine and\\ Integrated Frontier Research for Medical Science Division, \\Institute for Open and Transdisciplinary Research Initiatives (OTRI), The University of Osaka,\\
       Yamadaoka 2-2, Suita City, Osaka 565-0871, Japan\\
       \texttt{hattoris@biostat.med.osaka-u.ac.jp}\\
	   \And 
	   Tim Friede\\
       Department of Medical Statistics, University Medical Center G\"{o}ttingen and\\
       DZHK (German Center for Cardiovascular Research) and\\
       DZKJ (German Center for Child and Adolescent Health), \\
       Humboldtallee 32, 37073 G\"{o}ttingen, Germany\\
       \texttt{tim.friede@med.uni-goettingen.de}\\
       }

\begin{document}
\fontsize{12}{24}\selectfont

\maketitle

\begin{abstract}
For randomized controlled trials to be conclusive, it is important to set the target sample size accurately at the design stage. Comparing two normal populations, the sample size calculation requires specification of the variance other than the treatment effect and misspecification can lead to underpowered studies. Blinded sample size re-estimation is an approach to minimize the risk of inconclusive studies. Existing methods proposed to use the total (one-sample) variance that is estimable from blinded data without knowledge of the treatment allocation. We demonstrate that, since the expectation of this estimator is greater than or equal to the true variance, the one-sample variance approach can be regarded as providing an upper bound of the variance in blind reviews. This worst-case evaluation can likely reduce a risk of underpowered studies. However, blinded reviews of small sample size may still lead to underpowered studies. We propose a refined method accounting for estimation error in blind reviews using an upper confidence limit of the variance. A similar idea had been proposed in the setting of external pilot studies. Furthermore, we developed a method to select an appropriate confidence level so that the re-estimated sample size attains the target power. Numerical studies showed that our method works well and outperforms existing methods. The proposed procedure is motivated and illustrated by recent randomized clinical trials.
\end{abstract}


\section{Introduction}
In conducting clinical trials, the target sample size should be carefully determined to ensure scientific and ethical validity. A typical way to set the target sample size is to ensure the target power for the target treatment effect. Suppose we are conducting a clinical trial to compare two treatments regarding a continuous outcome. Since the power of the final analysis with the two-sample t-test relies on the common group-specific variance of the normal response across the two groups, specification of the variance is essential to calculate the sample size accurately. If there are some previously conducted clinical trials with similar designs, one may estimate the variance on the results of these studies \citep{schmidli2017meta}. On the other hand, there are often no or very limited information available and even if there are some information, they may not necessarily fit well to the target study. Thus, it is difficult to set the variance correctly at the design stage and the misspecification of the variance may lead to insufficient power of the final analysis. One may conduct some external pilot studies to obtain information on the variance. However, this is costly and, in particular, time-consuming.

Methods for sample size re-estimation based on mid-trial data review, which is also called internal pilot study, has been proposed to address this problem \citep{stein1945two}; they play an important role in new drug applications \citep{fda2019adaptive}. There are two types of sample size re-estimation methods based on variance estimation; unblinded and blinded ones. In the unblinded setting, the treatment allocation is revealed at the internal pilot study and then the treatment effects of the two groups and then the variance can be estimated by two-sample pooled variance estimator. \citet{wittes1999internal} showed that unblinded sample size re-estimation can ensure adequate power regardless of variance misspecification at the design stage. However, they also reported that (i) unblinded sample size re-calculation leads to inflation of the type I error rate and (ii) the sample size re-calculation may still lead to underpowered studies if the number of subjects enrolled in the internal pilot study is small, such as $10$ per group. To address the issue (i), several methods have been proposed to control the type I error rate \citep{denne1999estimating,kieser2000re,miller2005variance}. For the issue (ii), \citet{zucker1999internal} introduced the inflation factor approach to increase the re-estimated sample size empirically substituting the normal quantiles by quantiles of the t-distibution. Despite these methodological developments, unblinded sample size re-estimation requires access to treatment allocation information at the internal pilot study, which may raise concerns about operational bias. To minimize operational biases, it is preferable to conceal the allocation information at the internal pilot study \citep{fda2019adaptive}. Then, the blinded review is preferable in keeping integrity of clinical study. 

Under blinded review, available information for sample size re-calculation is restricted; the treatment effect of each group and then the variance cannot be estimated. Gould and Shih (1992) proposed use of the one-sample variance, which is the total variance of the pooled population estimatable with the blinded data. The one-sample variance has a positive bias for the common group-specific variance \citep{zucker1999internal} and then can be regarded as an upper bound of the variance. \citet{zucker1999internal} proposed a bias-adjusted one-sample variance estimator with the supposed target treatment effect. In the context of superiority trials, \citet{kieser2003simple} showed that blinded sample size re-calculation with those estimators does not lead to inflation of the type I error rate in contrast to unblinded sample size re-calculation. Also, \citet{kieser2003simple} compared blinded sample size re-calculation method with the one-sample variance and that of the bias-adjusted one-sample variance estimators and concluded that the unadjusted one-sample variance estimator is preferable for attaining the target power. In other word, taking a certain margin with the one-sample variance can be useful to protect underpowered studies. On the other hand, \citet{kieser2003simple} reported that small internal pilot study with the one-sample variance may still lead to underpowered studies. It implied that the margin put by the one-sample variance may not be enough and a more formal control of the margin would be beneficial. Against this background, based on the inflation factor approach originally proposed for unblinded sample size re-calculation \citep{zucker1999internal}, \citet{friede2013blinded} investigated its performance in blinded sample size re-estimation and concluded the inflation factor approach was also useful to avoid underpowered studies for blinded cases. However, its construction of the inflation factor was rather empirical and its theoretical basis is unclear.

We propose a new blinded sample size re-estimation method properly accounting for the estimation error at the internal pilot study. The idea of our method is to control the margin in estimation of the variance at the internal pilot study by using the upper confidence limit of the variance with a properly controlled confidence level. This idea was introduced by \citet{browne1995use}  for sample size calculation at the design stage with information from external pilot studies. \citet{browne1995use}  proposed to estimate the variance with its $100\cdot (1-\gamma)\%$ upper confidence limit for determining sample size. Based on the simulation results of \citet{browne1995use}, \citet{kieser1996use} investigated the theoretical properties of this approach and derived a method for selecting a relevant $1-\gamma$, depending on the number of subjects enrolled in the external pilot study, that controls the target power. In this paper, we extend this idea to the blinded internal pilot study setting. There are two difficulties to extend the method to our setting. Firstly, since the means of the two groups and then the variance cannot be consistently estimated under blind reviews, it is unclear how to construct the $100\cdot(1-\gamma)\%$ upper confidence limit of the variance tractable with blinded data.  Secondly, in the internal pilot studies, the data used for sample-size re-calculation are included in the final analysis, and the final sample size is dependent on part of the dataset for the final analysis. To handle these difficulties, we rely on asymptotic theory and use a conservative $100\cdot (1-\gamma)\%$ confidence
upper limit for the variance, which can be estimated in a blinded review. To select $1-\gamma$ attaining the target power of the final analysis, we derive a lower bound of the power based on the asymptotic conditional power given the re-estimated sample size \citep{zucker1999internal}. An interesting feature of our method is that given the number of subjects of the internal pilot study, the $1-\gamma$ can be determined independently of the variance. Thus, the choice of $1-\gamma$ can be pre-specified in study protocols. Through numerical studies, we found that our method can select a relevant $1-\gamma$ so as to avoid underpowered final analyses even with extremely small sample sizes of the internal pilot study. 

Our work is motivated by small-scale studies in clinical settings that are difficult to research. \citet{bannone2023acute} conducted a randomized clinical trial for the postpancreatectomy acute pancreatitis to evaluate whether the Diffusion-Weighted-MRI diagnosis work quickly or not. They conducted a blinded sample size re-calculation with $12$ subjects. \citet{duffley2023interactive} carried out a randomized clinical trial for Parkinson's disease to investigate whether Mobile Application for Pakinson's disease Deep brain stimulation can reduce programming time for the treatment of Pakinson's disease or not. They performed a blinded sample size re-calculation with $22$ subjects. These clinical trials suggest that some internal pilot studies are carried out with very small sample sizes; therefore there is a need for sample size re-estimation methods appropriate for small internal pilot studies. Our proposed method provides a relevant tool for this purpose. 

The remainder of the paper is organized as follows. In Section 2, we explain the settings we suppose and review existing methods for sample size re-calculation for a continuous outcome. In Section 3, we introduce our proposed method for blinded sample size re-calculation with a conservative upper confidence limit for the common group-specific variance and its theoretical rationale. In Section 4, we show the numerical study and results to evaluate the performance of the proposed method. In Section 5, we illustrate our proposed method with two real settings of clinical trials and conclude with a discussion in Section 6.

\section{Existing sample size re-estimation methods}
\subsection{Settings}
Suppose we are interested in designing a randomized trial comparing two treatments (experimental and control) regarding  a continuous outcome. Let $Z\in \{1,0\}$ be a treatment indicator with $P(Z=1)=\pi \hspace{1mm} (0<\pi <1)$; $Z=1$ if the subject is assigned to the experimental group and $Z=0$ if to the control group. Let $Y_z$ be the outcome following $N(\mu_{z},\sigma^2)$ if $Z=z$. Let $\Delta :=\mu_1-\mu_0$ be the true treatment effect and $\delta$ be a target treatment effect, which may be defined as the minimal clinically important difference. We set the target sample size, which has the target power $1-\beta$ (say 0.80) for the minimal clinically important difference $\delta$. Consider the following hypothesis testing problem;
\begin{align}
    H_0:\Delta=0\hspace{5mm}vs.\hspace{4mm}H_1:\Delta=\delta \hspace{1mm}(>0) \notag
\end{align}
We use Student's t-test with the one-sided significance level $\alpha$. Then, the target sample size of each group might be calculated using the following approximation;
\begin{align}
    n_{z,design}=P(Z=z) \Big(\frac{1}{\pi}+\frac{1}{1-\pi}\Big)\Big(\frac{z_\alpha - z_{1-\beta}}{\delta}\Big)^2 \sigma^2, \notag
\end{align}
where $z_{\alpha}$ is the upper $100\cdot \alpha\%$ quantile of the standard normal distribution \citep{chow2017sample}. To determine the sample size, in addition to the minimal clinically meaningful treatment effect $\delta$, we need to specify $\sigma^2$. In this paper, we call the parameter other than the treatment effect the nuisance parameter. Even if the experimental treatment actually has the expected treatment effect, inappropriate specification of $\sigma^2$ at the design stage may lead underpowered studies. To address this issue, sample size re-calculation at an internal pilot study is useful. In this paper, we take the blinded sample size re-estimation approach to ensure the target power. 

\subsection{A brief review of blinded sample size re-estimation methods for a normal outcome with an internal pilot study}
Let $n_{z,int}$ be the number of subjects included in the internal pilot study and $\hat{n}_{z,fin}$ be re-estimated sample size of the treatment group $Z=z$. Those for both groups are denoted by $n_{int}=n_{1,int}+n_{0,int}$ and $\hat{n}_{fin}=\hat{n}_{1,fin}+\hat{n}_{0,fin}$, respectively. The outcome $Y_z$ for the $i$th subject of the group $Z=z$ is denoted by $Y_{z,i}$. We suppose the first $n_{z,int}$ subjects are included in the internal pilot study without loss of generality. Under blinded review in which information of treatment assignments are not available, $\mu_z\hspace{0.5mm}(z\in\{1,0\})$ and then two-sample pooled variance cannot be calculated. \citet{gould1992sample} proposed to utilize the one-sample variance;
\begin{align}
    \hat{\sigma}^2_{OS}=\frac{1}{n_{int}-1} \sum_{z=0}^1\sum_{i=1}^{n_{z,int}}(Y_{z,i}-\bar{Y}_{int})^2 \label{OS-variance}
\end{align}
where $\bar{Y}_{int}=\sum_{z=0}^1 \sum_{i=1}^{n_{z,int}}Y_{z,i}/n_{int}$ is the sample mean of both groups at the internal pilot study. \citet{zucker1999internal} showed that the expectation of $\hat{\sigma}^2_{OS}$ is given by
\begin{align}
    E  [\hat{\sigma}^2_{OS}] =\sigma^2+\frac{n_{1,int}n_{0,int}}{n_{int}(n_{int}-1)} \Delta^2 \label{OS-expectation}
\end{align}
Then, $E[\hat{\sigma}^2_{OS}]\geq \sigma^2$. That is, $\hat{\sigma}^2_{OS}$ is an estimate of upper bound of $\sigma^2$ and then $\hat{\sigma}^2_{OS}$ is likely to overestimate $\sigma^2$. By setting $\Delta=\delta$, \citet{zucker1999internal} proposed the adjusted one-sample variance correcting the bias term of the one-sample variance;
\begin{align}
    \hat{\sigma}^2_{ADJ}=\hat{\sigma}^2_{OS}-\frac{n_{1,int}n_{0,int}}{n_{int}(n_{int}-1)} \delta^2     \notag
\end{align}
By \eqref{OS-expectation}, if $\Delta=\delta$ holds, $\hat{\sigma}^2_{ADJ}$ is an unbiased estimator of $\sigma^2$; $E[\hat{\sigma}^2_{ADJ}]=\sigma^2$. By estimating $\sigma^2$ with $\hat{\sigma}^2=\hat{\sigma}^2_{OS}$ or $\hat{\sigma}^2=\hat{\sigma}^2_{ADJ}$, the sample size of the group $Z=z$  is re-estimated by
\begin{align}
    \hat{n}_{z,fin} =P(Z=z) \Big(\frac{1}{\pi}+\frac{1}{1-\pi}\Big) \Big(\frac{z_\alpha - z_{1-\beta}}{\delta}\Big)^2 \hat{\sigma}^2  \hspace{1mm} \label{nfin}
\end{align}
and then the re-calculated total sample size is given by
\begin{align}
    \hat{n}_{fin} = \Big(\frac{1}{\pi}+\frac{1}{1-\pi}\Big) \Big(\frac{z_\alpha - z_{1-\beta}}{\delta}\Big)^2 \hat{\sigma}^2  \hspace{1mm}  \label{nfin_total}
\end{align}
After the sample size re-estimation, the final analysis is carried out once all outcomes of $\hat{n}_{fin}$ subjects have been observed. The test statistic $T$ is given by 
\begin{align}
    T=\frac{\bar{Y}_{1,fin}-\bar{Y}_{0,fin}}{\sqrt{\hat{\sigma}^2_{fin}(1/\hat{n}_{1,fin}+1/\hat{n}_{0,fin})}} \notag
\end{align}where $\bar{Y}_{z,fin}=\sum_{i=1}^{\hat{n}_{z,fin}}Y_{z,i}/\hat{n}_{z,fin}$ and $\hat{\sigma}^2_{fin}=\sum_{z=0}^1\sum_{i=1}^{\hat{n}_{z,fin}}(Y_{z,i}-\bar{Y}_{z,fin})^2/(\hat{n}_{fin}-2)$ are the sample means of each group and the two-sample pooled variance estimator at the final analysis, respectively. 

\citet{kieser2003simple} evaluated the power of the sample size re-calculation based on \eqref{nfin_total} with $\hat{\sigma}^2_{OS}$ and $\hat{\sigma}^2_{ADJ}$. Consider the conditional power $P(T\geq t_{\hat{n}_{fin}-2,\alpha}|\hat{n}_{fin},H_1)$ given $\hat{n}_{fin}$, where $t_{\hat{n}_{fin}-2,\alpha}$ is the upper $100\cdot\alpha\%$ quantile of the $t$ distribution with the degree of freedom $\hat{n}_{fin}-2$, evaluated by the exact (not asymptotic) distribution of the test statistic $T$. Since the sample size $\hat{n}_{fin}$ is dependent on data at the internal pilot study, the power was derived by marginalizing the conditional power given $\hat{n}_{fin}$, $P(T\geq t_{\hat{n}_{fin}-2,\alpha}|\hat{n}_{fin},H_1)$, by the distribution of $\hat{n}_{fin}$. They concluded that $\hat{\sigma}^2_{ADJ}$ that is unbiased under the specific alternative  might lead to  underpowered final analyses and recommended the use of $\hat{\sigma}^2_{OS}$ to avoid inconclusive studies. However, as shown in Table 2 of \citet{kieser2003simple}, the sample size re-estimation method using $\hat{\sigma}^2_{OS}$ does not necessarily maintain the target power when $n_{int}$ is small. To address this issue, \citet{friede2013blinded} applied the inflation factor approach \citep{zucker1999internal} to blinded sample size re-estimation and investigated the power with the re-estimated sample size \eqref{nfin_total} increased by the inflation factor, $IF=(t_{n_{int}-2,\alpha}-t_{n_{int}-2,1-\beta})^2/(z_{\alpha}-z_{1-\beta})^2$, with $\hat{\sigma}^2=\hat{\sigma}^2_{OS}$;
\begin{align}
    \hat{n}_{fin}=\Big(\frac{1}{\pi}+\frac{1}{1-\pi}\Big)\Big(\frac{t_{n_{int}-2,\alpha}-t_{n_{int}-2,1-\beta}}{\delta}\Big)^2\hat{\sigma}^2_{OS} \label{nfin_zucker}
\end{align}

\section{Blinded sample size re-estimation with upper confidence limit accounting for uncertainty in mid-trial estimation}
\subsection{The motivation and idea of the proposed method}
As discussed in Section 2.2, use of $\hat{\sigma}^2_{OS}$ is regarded as taking a certain margin in estimation of $\sigma^2$ with an upper bound subject to blinded data of the internal pilot study. However, it is unclear why this upper bound should be used and more proper management of overestimation may lead to more favourable properties of the procedure in terms of power and sample size distribution. In this paper, we consider controlling the power by utilizing confidence intervals at the internal pilot studies. That is, we consider the sample size re-estimation with the upper confidence limit for the variance constructed using $\hat{\sigma}^2_{OS}$ under blinded review and control the power selecting a proper confidence level, $100\cdot (1-\gamma)\%$ attaining the target power. In the context of sample size calculation at the design stage, there are some existing methods accounting for the upper confidence limit estimated at the external pilot study \citep{browne1995use,kieser1996use}. \citet{kieser1996use} showed that $1-\gamma$ can be determined so that the power equals to the target power, since the power is expressed with only $1-\gamma$ and the number of subjects in the external pilot study given the significance level of the final statistical analysis. In our study, we apply their idea to internal pilot studies for blinded sample size re-estimation. There are two points to consider. First, since \citet{kieser1996use} handled the external pilot study, parameters estimated with the external pilot study for sample size re-estimation are statistically independent of the final statistics. Second, \citet{kieser1996use} focused on the unblinded case and then one can estimate group-specific parameters and then can estimate the upper confidence limit of $\sigma^2$. That is the reason why their power formula can be expressed without the effect size $\Delta/\sigma$  and then determine $1-\gamma$ attaining the target power of the final analysis. In the context of the blinded sample size re-calculation, \citet{kieser2003simple} calculated the power of the t-test with its exact (not asymptotic) distribution. In addition to $1-\gamma$ and $n_{int}$, their power  is dependent on the effect size $\Delta/\sigma$. Note that the effect size is unknown and then \citet{kieser2003simple} utilized the formula for the power only for evaluating performance of the sample size re-calculation methods and did not do for determining the sample size. Furthermore, the upper confidence limit for $\sigma^2$ constructed using $\hat{\sigma}^2_{OS}$ is also dependent on the non-centrality parameter $\lambda$ of the non-central chi-squared distribution, which depends on the effect size $\Delta/\sigma$ (see Section 3.2). Therefore, the upper confidence limit is not estimable in a blinded review and it is a challenging task to apply the idea by \citet{kieser1996use} to use the $100\cdot(1-\gamma)\%$ upper confidence limit for the power. A new idea is needed to overcome these difficulties, and we propose to use a conservative upper confidence limit, which can be constructed under a blinded review, and derive the lower bound of the power that does not depend on the effect size $\Delta/\sigma$. By setting $1-\gamma$ attaining the lower bound of the power to the target power, one may avoid underpowered studies.

\subsection{Asymptotically valid but infeasible upper confidence limit approach}
Firstly, we show how to construct asymptotically valid confidence interval for $\sigma^2$ with an internal pilot study and the formula to calculate the power based on the upper limit of the confidence interval. The formula is shown to be dependent on $\Delta/\sigma$. Then it cannot be applied under a blinded review. We still consider it here to clarify some aspects of our idea although practically infeasible. Let $W$ be a random variable which follows to $\chi^2_{n_{int}-1}$, the central chi-squared distribution with $n_{int}-1$ degrees of freedom, and let $W_{\lambda}$ be a random variable which follows to $\chi^2_{n_{int}-1,\lambda}$, the non-central chi-squared distribution with $n_{int}-1$ degrees of freedom and non-centrality parameter $\lambda$. 
Denote $F_W$, $f_W$ and $d_{1-\gamma}$ be the cumulative distribution function, the probability density function and the upper $100\cdot \alpha\%$ quantile of $\chi^2_{n_{int}-1}$, respectively.
Those for the non-central chi-squared distribution $\chi^2_{n_{int-1},\lambda}$ are denoted by $F_{W_\lambda}$, $f_{W_\lambda}$ and $d_{\lambda,1-\gamma}$, respectively. 
Note that $\hat{\sigma}^2_{OS}(n_{int}-1)/\sigma^2\sim \chi^2_{n_{int}-1,\lambda}$ with the non-centrality parameter $\lambda = (\Delta/\sigma)^2 (n_{1,int}n_{0,int})/n_{int}$. Then, it holds that  
\begin{align}
     P\Big(\frac{\hat{\sigma}^2_{OS}(n_{int}-1)}{d_{\lambda,1-\gamma}}\geq\sigma^2\Big)=P\Big(\frac{\hat{\sigma}^2_{OS}(n_{int}-1)}{\sigma^2}\geq d_{\lambda, 1-\gamma}\Big)=P(W_{\lambda}\geq d_{\lambda,1-\gamma})=1-\gamma . \notag
\end{align}
Therefore, a $100\cdot(1-\gamma)\%$ upper confidence limit for $\sigma^2$ is given by
\begin{align}
    \hat{\sigma}^2_{U,1-\gamma}=\hat{\sigma}^2_{OS} \frac{n_{int}-1}{d_{\lambda,1-\gamma}} \label{accu_CI}
\end{align} 
Since $\hat{\sigma}^2_{U,1-\gamma }$ is dependent on $d_{\lambda,1-\gamma}$ and the non-centrality parameter $\lambda$ depends on the unknown effect size $\Delta/\sigma$, it is not estimable under a blinded review. We present the power formula based on the two-sample variance estimator $\hat{\sigma}^2_{fin}$ as follows, which is infeasible in a blinded condition, but useful for our development. Since $\hat{\sigma}^2_{fin}$ included in the test statistic $T$ and $t_{\hat{n}_{fin}-2,\alpha}$ asymptotically approximate $\sigma^2$ and $z_{\alpha}$, respectively, the asymptotic conditional power given $\hat{n}_{fin}$ is expressed as follows \citep{zucker1999internal};
\begin{align}
P(T\geq t_{\hat{n}_{fin}-2,\alpha}|\hat{n}_{fin},H_1)&=
 P\Big(\frac{\bar{Y}_{1,fin}-\bar{Y}_{0,fin}}{\sqrt{\hat{\sigma}^2_{fin}(1/\hat{n}_{1,fin}+1/\hat{n}_{0,fin})}}\geq t_{\hat{n}_{fin}-2,\alpha}|\hat{n}_{fin},H_1\Big)\notag\\
& \approx P\Big(\frac{\bar{Y}_{1,fin}-\bar{Y}_{0,fin}}{\sqrt{\sigma^2(1/\hat{n}_{1,fin}+1/\hat{n}_{0,fin})}}\geq z_{\alpha}|\hat{n}_{fin},H_1\Big)\notag\\&=
    1-\Phi\Big(z_{\alpha}-\frac{\Delta}{\sqrt{\sigma^2\{1/\pi+1/(1-\pi)\}/\hat{n}_{fin}}}\Big) \label{zucker2}
\end{align}
Let us consider to calculate $\hat{n}_{fin}$ with $\hat{\sigma}^2_{U,1-\gamma}$ based on \eqref{nfin};
\begin{align}
    \hat{n}_{fin}=\Big(\frac{1}{\pi}+\frac{1}{1-\pi}\Big)\Big(\frac{z_{\alpha}-z_{1-\beta}}{\delta}\Big)^2\hat{\sigma}^2_{U,1-\gamma} \label{nfin_accu}
\end{align}
Substituting $\hat{n}_{fin}$ in \eqref{zucker2} by \eqref{nfin_accu} results in
\begin{align}
 P(T\geq t_{\hat{n}_{fin}-2,\alpha}|\hat{n}_{fin},H_1) \approx 
    1-\Phi\Big(z_{\alpha}-\frac{\Delta}{\delta}\cdot|z_{\alpha}-z_{1-\beta}|\cdot\sqrt{\frac{\hat{\sigma}^2_{U,1-\gamma}}{\sigma^2}}\Big) \label{blind_cpower}
\end{align}
Since $\hat{\sigma}^2_{U,1-\gamma}/\sigma^2=W_{\lambda}/d_{\lambda,1-\gamma}$ and \eqref{blind_cpower}, the power is expressed as
\begin{align}
 P(T\geq t_{\hat{n}_{fin}-2,\alpha}|H_1) \approx 
    1-\int_{0}^\infty\Phi\Big(z_{\alpha}-\frac{\Delta}{\delta}\cdot |z_{\alpha}-z_{1-\beta}|\cdot\sqrt{\frac{w}{d_{\lambda,1-\gamma}}}\Big)f_{W_\lambda}(w)dw  \label{blind_epower}
\end{align}
The right hand side of \eqref{blind_epower} is expressed by $d_{\lambda,1-\gamma}$ and $W_{\lambda}$. Since $\lambda$ includes the unknown effect size $\Delta/\sigma$, it is not possible to calculate \eqref{accu_CI} and \eqref{blind_epower} in practice.

\subsection{Proposed method: Use of a conservative upper confidence limit of $\sigma^2$ estimable in a blind review}
The notation is summarized in Table 1. In the following, we use some properties on the relationship between the central and non-central chi-squared distributions with common degree of freedom. We begin by preparing the following lemmas.
\vskip\baselineskip
\begin{lemma}
For the central and non-central chi-squared distributions with common degree of freedom, the following inequality holds:
\begin{align}
    F_{W_{\lambda}}(w)\leq F_W (w) \hspace{1mm}(\forall w\geq 0) \notag
\end{align}
Hence, the following inequality also holds:
\begin{align}
    d_{\lambda,1-\gamma}\geq d_{1-\gamma} \notag
\end{align}
\end{lemma}
\vskip\baselineskip
\begin{lemma}
    Let $F_1,F_2$ be distribution functions of $X_1,X_2$, that satisfies $F_1(x)\leq F_2(x)\hspace{1mm}(\forall x\in\mathbb{R})$. Suppose a non-decreasing function $g$ satisfies $E[g(X_1)]<\infty$ and $E[g(X_2)]<\infty$, then the following inequality holds:
    \begin{align}
        E[g(X_1)]\geq E[g(X_2)] \notag
    \end{align}
\end{lemma}
\vskip\baselineskip
\noindent
The proofs of Lemma 1 and 2 are given by \citet{finner1997log} and \citet[pp.~757--777]{lehmann1955ordered}, respectively.
Define $\bar{\sigma}^2_{U,1-\gamma}=\hat{\sigma}^2_{U,1-\gamma}d_{\lambda,1-\gamma}/d_{1-\gamma}$. Since \eqref{accu_CI}, 
\begin{align}
    \bar{\sigma}^2_{U,1-\gamma}=\hat{\sigma}^2_{U,1-\gamma}\frac{d_{\lambda,1-\gamma}}{d_{1-\gamma}}=\hat{\sigma}^2_{OS}\frac{n_{int}-1}{d_{1-\gamma}} \label{cons_ci}
\end{align}
holds. Note that it does not depend on $\Delta/\sigma$. From \eqref{cons_ci}, one has $\bar{\sigma}^2_{U,1-\gamma}d_{1-\gamma}=\hat{\sigma}^2_{OS}(n_{int}-1)$, then $\bar{\sigma}^2_{U,1-\gamma}d_{1-\gamma}/\sigma^2$ follows the non-central chi-squared distribution $W_{\lambda}$ since $\hat{\sigma}^2_{OS}(n_{int}-1)/\sigma^2\sim W_{\lambda}$. That is $\bar{\sigma}^2_{U,1-\gamma}/\sigma^2=W_{\lambda}/d_{1-\gamma}$ holds. This is a key quantity in our development. It holds that
\begin{align}
    P(\bar{\sigma}^2_{U,1-\gamma}\geq \sigma^2)=P\Big(\hat{\sigma}^2_{U,1-\gamma}\cdot \frac{d_{\lambda,1-\gamma}}{d_{1-\gamma}}\geq \sigma^2\Big)\geq P(\hat{\sigma}^2_{U,1-\gamma}\geq \sigma^2)=1-\gamma \notag
\end{align}
The inequality follows from Lemma 1. 
\noindent
Thus, $\bar{\sigma}^2_{U,1-\gamma}$ gives us a conservative $100\cdot (1-\gamma)\%$ upper confidence limit for $\sigma^2$. 
Let us calculate $\hat{n}_{fin}$ with $\bar{\sigma}^2_{U,1-\gamma}$ based on \eqref{nfin};
\begin{align}
    \hat{n}_{fin}=\Big(\frac{1}{\pi}+\frac{1}{1-\pi}\Big)\Big(\frac{z_{\alpha}-z_{1-\beta}}{\delta}\Big)^2\bar{\sigma}^2_{U,1-\gamma}   \label{nfin_cons}
\end{align}
Substituting $\hat{n}_{fin}$ in \eqref{zucker2} by \eqref{nfin_cons}, 
\begin{align}
P(T\geq t_{\hat{n}_{fin}-2,\alpha}|\hat{n}_{fin},H_1) \approx 
    1-\Phi\Big(z_{\alpha}-\frac{\Delta}{\delta}\cdot |z_{\alpha}-z_{1-\beta}|\cdot \sqrt{\frac{\bar{\sigma}^2_{U,1-\gamma}}{\sigma^2}}\Big) \label{blind_cpower_2}
\end{align}
holds. With $\bar{\sigma}^2_{U,1-\gamma}/\sigma^2=W_{\lambda}/d_{1-\gamma}$ and \eqref{blind_cpower_2}, 
the power is given by
\begin{align}
P(T\geq t_{\hat{n}_{fin}-2,\alpha}|H_1) \approx 
    1-\int_{0}^\infty\Phi\Big(z_{\alpha}-\frac{\Delta}{\delta}\cdot |z_{\alpha}-z_{1-\beta}|\cdot\sqrt{\frac{w}{d_{1-\gamma}}}\Big)f_{W_\lambda}(w)dw \label{blind_epower_2}
\end{align}
Formula \eqref{blind_epower_2} still includes the non-centrality parameter $\lambda=(\Delta/\sigma)^2(n_{1,int}n_{0,int})/n_{int}$, which depends on the effect size $\Delta/\sigma$. Then, we consider a lower bound of \eqref{blind_epower_2}. 
We apply Lemma 2 to the distributions of $W$ and $W_{\lambda}$. Let $g(w)$ be the integrand of \eqref{blind_epower_2};
\begin{align}
    g(w)=1-\Phi\Big(z_{\alpha}-\frac{\Delta}{\delta}\cdot|z_{\alpha}-z_{1-\beta}|\cdot \sqrt{\frac{w}{d_{1-\gamma}}}\Big)\notag
\end{align}
Here, $g(w)$ is a non-decreasing function under $H_1:\Delta=\delta$. Although $g(w)$ is not a non-decreasing function in general, it is sufficient to consider $g(w)$ under $H_1$ to evaluate the lower bound of the power. 
The power \eqref{blind_epower_2} is represented as $E[g(W_{\lambda})]$. From Lemma 1, $F_{W_{\lambda}}(w)\leq F_W (w)\hspace{0.5mm}(\forall w\geq 0)$ holds. Then, from Lemma 2,  
\begin{align}
    E[g(W_{\lambda})]&\geq E[g(W)]\notag\\
    &=1-\int_0^{\infty} \Phi\Big(z_{\alpha}-\frac{\Delta}{\delta}\cdot|z_{\alpha}-z_{1-\beta}|\cdot \sqrt{\frac{w}{d_{1-\gamma}}}\Big)f_{W}(w) \label{lowerbound}
\end{align}
holds. It gives a lower bound of the power $E[g(W_\lambda)]$ in \eqref{blind_epower_2} and is free of $\lambda$ and $\sigma^2$. Under $H_1:\Delta=\delta$, given $n_{int}$ and the significance level $\alpha$ of the final analysis, one can calculate \eqref{lowerbound} depending only on $1-\gamma$ and then determine $1-\gamma$ so that \eqref{lowerbound} attains the target power. With the resulting $1-\gamma$, one can determine $\hat{n}_{fin}$ with \eqref{nfin_cons} ensuring the target power. In summary, we propose the following procedure; 
\begin{enumerate}[label=Step \arabic*.]
    \item (Design stage) Set the one-sided significance level $\alpha$, the target power $1-\beta$, the target treatment effect $\delta$ and the number of sample size of the blinded sample size re-calculation $n_{int}$. 
    \item (Design stage) Determine $1-\gamma$ satisfying \eqref{lowerbound} $=1-\beta$ and define it in the study protocol. 
    \item (Internal pilot study) Calculate $\bar{\sigma}^2_{U,1-\gamma}$ and then determine $\hat{n}_{fin}$ with \eqref{nfin_cons}.
    \item (Final analysis) Carry out the final analysis with $\hat{n}_{fin}$ subjects.
\end{enumerate}

 An interesting feature of the proposed procedure is that $1-\gamma$ can be determined independently of data. In Table 2, $1-\gamma$ is listed for some settings of $n_{int}$. One may present a similar table for some plausible range of $n_{int}$ in the study protocol.

\section{Numerical studies}
In this section, we evaluate performance of the proposed method. We ask (i) whether the proposed method well control the power and then contribute avoiding underpowered studies, (ii) whether the proposed blinded sample size re-calculation method does not lead to inflation of the type I error rate, and (iii) what distribution the re-calculated sample size follows. For (i) and (ii), we evaluated theoretical values of the type I error rate and the power by \citet{kieser2003simple}, see line $11$, page $6$ and line $13$, page $5$ of \citet{kieser2003simple} for the formula. To calculate the integral in the formula, instead of numerical integration, we used Monte-Carlo integration with $10^8$ uniform random numbers generated. We begin with evaluating the power to address the issue (i), following the setting of \citet{kieser2003simple} as follows. Suppose we are conducting a randomized clinical trial with a continuous outcome. Subjects are randomly allocated to either of the treatment and the control groups with the allocation rate $\pi=0.5$. We set the target sample size for the t-test with one-sided significance level of $\alpha=0.025$, the target power $1-\beta=0.80$ and the target treatment effect $\delta=1$. We consider three setting of the variance $\sigma^2=2.038$, $4.013$ and $11.08$. To attain the target power, one needs to enroll $N=32$, $63$ and $174$ subjects per group, respectively. Blinded sample size re-calculation was conducted. We applied the following methods; the method by Gould and Shih (1992) with the one-sample variance in \eqref{nfin_total} referred as {\it One-sample}, the inflation factor approach by \citet{zucker1999internal} in \eqref{nfin_zucker} as {\it IF} and the proposed method with the variance estimate $\bar{\sigma}^2_{U,1-\gamma}$ in \eqref{nfin_cons} as {\it Proposed}. We also applied the method in Section 3.2 with $\hat{\sigma}^2_{U,1-\gamma}$ in \eqref{nfin_accu} for reference, which is referred as {\it Theoretical}. In practice, the number of sample size at blinded reviews would not be equal, even if the allocation rate is $\pi=0.5$. \citet{kieser2003simple} calculated the power under the assumption $n_{1,int}=n_{0,int}$. In contrast, we derived the power without this assumption and carried out the numerical studies. However, we only show the results with $n_{1,int}=n_{0,int}$ since the results were similar with unequal sample sizes. In Table 3, we demonstrate the power. Consistently with \citet{kieser2003simple}, the power with the {\it One-sample} method could be less than the target power of $0.80$ if the sample size of the internal pilot study was extremely small. For example, with the internal pilot study with $10$ subjects ($n_{z,int}=5$), the power with the {\it One-sample} method was $0.7517$ for $N=32$. On the other hand, the proposed method controlled the power very well. For example, with $10$ subjects in the internal pilot study, the {\it Proposed} method had the power $0.8153$ for $N=32$. The {\it IF} method by \citet{zucker1999internal} could avoid underpowered studies as seen in Table 3. However, magnitude of overestimation of the power was larger with the {\it IF} method than with the {\it Proposed} method with $N=32$ and $63$. The {\it Proposed} method successfully controlled the power regardless of the size of the study. Interestingly, the {\it Theoretical} method could not achieve the target power. Recall that the {\it Proposed} method takes a further margin than the {\it Theoretical} method considering a conservative power evaluation by \eqref{lowerbound}. It led more accurate re-calculated sample size. 

Next, we evaluate the type I error rates. Applying the formula by \citet{kieser2003simple} with $\Delta=0$, the type I error rates of the final analysis with a re-calculated sample size was evaluated for $N=32$, $63$ and $174$ in Table 4. The type I error rates were not substantially inflated for any of the methods. 

Finally, we evaluate the variation of the re-calculated sample size for the final analysis. Since $\hat{n}_{z,fin}$ is a function of $\hat{\sigma}^2_{OS}$ as given in \eqref{nfin_total} and the distribution of $\hat{\sigma}^2_{OS}$ is given by the non-central chi-squared distribution, the re-calculated sample size $\hat{n}_{z,fin}$ with the {\it One-sample} method can be investigated by generating $\hat{\sigma}^2_{OS}$ in computer (see line 6, page 5 of \citet{kieser2003simple}). Similarly, the re-calculated sample size can be generated with \eqref{nfin_zucker}, \eqref{nfin_accu} and \eqref{nfin_cons} for the {\it IF}, the {\it Theoretical} and the {\it Proposed} methods, respectively. Averages and empirical standard deviations of the $10^8$ generated $\hat{n}_{z,fin}$ are summarized in Table 5 for the {\it One-sample}, the {\it IF}, the {\it Proposed} and the {\it Theoretical} methods. In particular, the distributions of $\hat{n}_{z,fin}$ with the {\it One-sample} method and the {\it Proposed} method are compared in Figure 1, showing the median, the $1^{st}$ and $3^{rd}$ quartiles. They indicate that the re-calculated sample size $\hat{n}_{z,fin}$ was highly variable with the internal pilot studies of extremely small sample size such as $4$ in total. As argued, even with $4$ subjects in the internal pilot study, the power with the {\it Proposed} method was close to the target power and then could control the power well. However, we must pay cost of high variation of the re-calculated sample size with extremely small internal pilot studies.

\section{Clinical trial examples}
\subsection{A randomized clinical trial for postpancreatectomy acute pancreatitis}
\citet{bannone2023acute} investigated whether the Diffusion-Weighted-MRI (DW-MRI) diagnosis worked quickly or not for the postpancreatectomy acute pancreatitis (PPAP) and postoperative hyperamylasemia (POH). The primary outcome was the pancreatic apparent diffusion coefficient (ADC) value at DW-MRI and  its mean was compared between patients with PPAP and POH and those without PPAP or POH. \citet{bannone2023acute} determined sample size $n_{design}=24$ under the following  assumptions; $\alpha=0.05$ (two-sided), $1-\beta=0.85$, $\Delta=\delta=4.5\times 10^{-4}, \sigma^2=1.35\times 10^{-7}$. \citet{bannone2023acute} applied the sample size re-estimation with the one-sample variance $\hat{\sigma}^2_{OS}$ at $n_{int}=12$ and sample size was re-estimated as $\hat{n}_{fin}=65$. At the final analysis, the treatment effect was estimated as $\hat{\Delta}=2.3\times 10^{-4}$ and p-value was calculated as $0.006$. 

Under this setting, we applied our proposed method calculating \eqref{nfin_cons}. To conduct it, we assumed the outcome followed the normal distributions and a balanced allocation ($n_{1,int}=n_{0,int},\pi=0.5$). Although \citet{bannone2023acute} did not report $\hat{\sigma}^2_{OS}$, we calculated it from the result of their sample size re-stimation as $\hat{\sigma}^2_{OS}=3.67\times 10^{-7}$. Given $n_{int}=12$, we calculate $1-\gamma=0.62$ that satisfies \eqref{lowerbound} equals to the target power $1-\beta=0.85$. Finally, under the settings of $\hat{\sigma}^2_{OS}=3.67\times 10^{-7}$, $1-\gamma =0.62$, $\alpha =0.025$ (one-sided), $1-\beta=0.85$, we have $\bar{\sigma}^2_{U,1-\gamma}=\bar{\sigma}^2_{U,0.62}=4.48\times 10^{-7}$ and then calculate $\hat{n}_{fin}=80$ with \eqref{nfin_cons}. Thus, although \citet{bannone2023acute} successfully obtained a statistically significant result ($p=0.006$), the one-sample variance $\hat{\sigma}^2_{OS}$ might have been too small to ensure the taret power. This is consistent with the observations in the numerical study. 

\subsection{A randomized clinical trial for Parkinson's disease}
\citet{duffley2023interactive} investigated whether Mobile Application for Parkinson's disease Deep brain stimulation (MAP DBS) can reduce programming time for Deep brain stimulation (DBS), which is an effective treatment for Parkinson's disease (PD). As the primary outcome, the mean total programming time difference between the standard of care DBS (SOC DBS) group and the MAP DBS group was estimated using a log transformation. \citet{duffley2023interactive} determined sample size $n_{design}=72$ under the following assumptions; $\alpha=0.05$ (two-sided), $1-\beta=0.80$, $\Delta=\delta=0.40, \sigma^2=0.362$. \citet{duffley2023interactive} conducted a sample size re-calculation with the one-sample variance $\hat{\sigma}^2_{OS}=0.192$ at $n_{int}=22$. The paper by \citet{duffley2023interactive} did not report the resulting re-calculated sample size, which was re-constructed by our calculation as $n_{int}=38$. Finally, \citet{duffley2023interactive} did not modify the sample size and conducted the final analysis with $n_{design}=72$ reporting the treatment effect as $\hat{\Delta}=0.47$ with p-value of $0.001$. 

Under this setting, we applied our proposed method calculating \eqref{nfin_cons}. To conduct it, we assumed the outcome followed the log-normal distributions and a balanced allocation ($n_{1,int}=n_{0,int},\pi=0.5$). Given $n_{int}=22$, we calculated $1-\gamma=0.57$ that satisfies \eqref{lowerbound} equals to the target power $1-\beta=0.80$. Finally, under the settings of $\hat{\sigma}^2_{OS}=0.192$, $1-\gamma =0.57$, $\alpha =0.025$ (one-sided), $1-\beta=0.80$, we obtained $\bar{\sigma}^2_{U,1-\gamma}=\bar{\sigma}^2_{U,0.57}=0.210$ and then calculate $\hat{n}_{fin}=42$ with \eqref{nfin_cons}. 
Thus, if \citet{duffley2023interactive} modified the sample size from the original $72$ to $38$ instead of $42$, it may result in a lack of power in the final analysis.

\section{Discussion}
In this paper, we proposed a blinded sample size re-calculation method. Due to estimation errors at the internal pilot study, unbiased estimators for nuisance parameters associated with the power might lead to underpowered studies. Instead, taking a certain margin in estimation would improve the performance of the sample size. Our method provides more formal way to take a margin than the widely used one-sample variance approach. It successfully overcomes a drawback of the one-sample variance approach; our method can control the power even if the sample size of the internal pilot study is less than $20$ and surprisingly maintain the target power with the internal pilot study of extremely small sample size like $n_{int}=4$. 

However, we need to re-emphasize that with extremely small internal pilot studies we must pay a cost of large variation of the re-calculated sample size and then sample size re-calculation may suggest unrealistically large sample sizes (e.g., blinded sample size re-calculation with subgroup analyses; \citet{placzek2023blinded}). Typically, sample size re-calculation is used for confirmatory clinical trials with certainly large number of subjects and then is usually conducted with moderate or large sample size. However, as shown in Section $5$, some clinical trials are conducted with extremely small internal pilot studies. It should be useful for the clinical trials of rare disease that are difficult to recruit patients, from the perspective of avoiding the risk of the trial failure due to the underpowered study, while the issue of large variability of the re-calculated sample size $\hat{n}_{fin}$ should be appropriately addressed. 

The success of our proposed method was to introduce conservative evaluation of the power. The key technical tool for evaluation came from relationships between the central and the non-central chi-squared distributions summarized as Lemma $1$ and Lemma $2$ in Section $3$. Although they were established long ago \citep{lehmann1955ordered,finner1997log}, to our best knowledge, they have not been applied in developing adaptive design methods. We expect the techniques used in this paper would be useful for other problems of adaptive designs. 

Although we focus on blinded sample size re-calculation in this paper, our theory is also applicable to the unblinded setting. Consider using the upper confidence limit constructed from two-sample pooled variance estimator (provided e.g. in \citet{kieser1996use}) \color{black}that is estimable in an unblinded review. This procedure was considered by \citet{kieser2000re} for various confidence levels $1-\gamma$ ranging from $60-90\%$ with control of the Type I error rate. Based on the procedure proposed here, \color{black}one could extend their approach by \color{black}deriving the power depending only on $1-\gamma$ and determine $1-\gamma$ so as to achieve the target power. With the resulting $1-\gamma$, one can re-estimate the sample size ensuring the target power even with a small internal pilot study.

\bibliographystyle{apalike}
\bibliography{reference}

@article{bannone2023acute,
  title={Acute pancreatitis after pancreatoduodenectomy: A prospective study of diffusion-weighted magnetic resonance imaging, serum biomarkers, and clinical features},
  author={Bannone, Elisa and Marchegiani, Giovanni and Zamboni, Giulia Angela and Maris, Bogdan Mihai and Costa, Lorenzo and Procida, Giuseppa and Vacca, Pier Giuseppe and D’Onofrio, Mirko and Mansueto, Giancarlo and De-Madaria, Enrique and others},
  journal={Surgery},
  volume={173},
  number={6},
  pages={1428--1437},
  year={2023},
  publisher={Elsevier}
}

@article{browne1995use,
  title={On the use of a pilot sample for sample size determination},
  author={Browne, Richard H},
  journal={Statistics in Medicine},
  volume={14},
  number={17},
  pages={1933--1940},
  year={1995},
  publisher={Wiley Online Library}
}

@book{chow2017sample,
  title={Sample size calculations in clinical research},
  author={Chow, Shein-Chung and Shao, Jun and Wang, Hansheng and Lokhnygina, Yuliya},
  year={2017},
  publisher={chapman and hall/CRC}
}

@article{denne1999estimating,
  title={Estimating the sample size for at-test using an internal pilot},
  author={Denne, Jonathan S and Jennison, Christopher},
  journal={Statistics in Medicine},
  volume={18},
  number={13},
  pages={1575--1585},
  year={1999},
  publisher={Wiley Online Library}
}

@article{duffley2023interactive,
  title={Interactive mobile application for Parkinson's disease deep brain stimulation (MAP DBS): An open-label, multicenter, randomized, controlled clinical trial},
  author={Duffley, Gordon and Szabo, Aniko and Lutz, Barbara J and Mahoney-Rafferty, Emily C and Hess, Christopher W and Ramirez-Zamora, Adolfo and Zeilman, Pamela and Foote, Kelly D and Chiu, Shannon and Pourfar, Michael H and others},
  journal={Parkinsonism \& related disorders},
  volume={109},
  pages={105346},
  year={2023},
  publisher={Elsevier}
}

@article{finner1997log,
  title={Log-concavity and inequalities for Chi-square, F and Beta distributions with applications in multiple comparisons},
  author={Finner, Helmut and Roters, Markus},
  journal={Statistica Sinica},
  volume={7},
  number={3},
  pages={771--787},
  year={1997},
  publisher={JSTOR}
}

@misc{fda2019adaptive,
  author       = {{Food and Drug Administration (FDA)}},
  title        = {Adaptive Design Clinical Trials for Drugs and Biologics: Guidance for Industry},
  year         = {2019},
  howpublished = {\url{https://www.fda.gov/regulatory-information/search-fda-guidance-documents/adaptive-design-clinical-trials-drugs-and-biologics-guidance-industry}}
}

@article{friede2013blinded,
  title={Blinded sample size re-estimation in superiority and noninferiority trials: bias versus variance in variance estimation},
  author={Friede, Tim and Kieser, Meinhard},
  journal={Pharmaceutical Statistics},
  volume={12},
  number={3},
  pages={141--146},
  year={2013},
  publisher={Wiley Online Library}
}

@article{gould1992sample,
  title={Sample size re-estimation without unblinding for normally distributed outcomes with unknown variance},
  author={Gould, A Lawrence and Shih, Weichung Joseph},
  journal={Communications in Statistics-Theory and Methods},
  volume={21},
  number={10},
  pages={2833--2853},
  year={1992},
  publisher={Taylor \& Francis}
}

@article{kieser2000re,
  title={Re-calculating the sample size in internal pilot study designs with control of the type I error rate},
  author={Kieser, Meinhard and Friede, Tim},
  journal={Statistics in Medicine},
  volume={19},
  number={7},
  pages={901--911},
  year={2000},
  publisher={Wiley Online Library}
}

@article{kieser2003simple,
  title={Simple procedures for blinded sample size adjustment that do not affect the type I error rate},
  author={Kieser, Meinhard and Friede, Tim},
  journal={Statistics in Medicine},
  volume={22},
  number={23},
  pages={3571--3581},
  year={2003},
  publisher={Wiley Online Library}
}

@article{kieser1996use,
  title={On the use of the upper confidence limit for the variance from a pilot sample for sample size determination},
  author={Kieser, Meinhard and Wassmer, Gernot},
  journal={Biometrical Journal},
  volume={38},
  number={8},
  pages={941--949},
  year={1996},
  publisher={Wiley Online Library}
}

@incollection{lehmann1955ordered,
  title={Ordered families of distributions},
  author={Lehmann, Erich Leo},
  booktitle={Selected Works of EL Lehmann},
  pages={757--777},
  year={1955},
  publisher={Springer}
}

@article{miller2005variance,
  title={Variance estimation in clinical studies with interim sample size reestimation},
  author={Miller, Frank},
  journal={Biometrics},
  volume={61},
  number={2},
  pages={355--361},
  year={2005},
  publisher={Oxford University Press}
}

@article{placzek2023blinded,
  title={Blinded sample size recalculation in adaptive enrichment designs},
  author={Placzek, Marius and Friede, Tim},
  journal={Biometrical Journal},
  volume={65},
  number={2},
  pages={2000345},
  year={2023},
  publisher={Wiley Online Library}
}

@article{schmidli2017meta,
  title={Meta-analytic-predictive use of historical variance data for the design and analysis of clinical trials},
  author={Schmidli, Heinz and Neuenschwander, Beat and Friede, Tim},
  journal={Computational Statistics \& Data Analysis},
  volume={113},
  pages={100--110},
  year={2017},
  publisher={Elsevier}
}

@article{stein1945two,
  title={A two-sample test for a linear hypothesis whose power is independent of the variance},
  author={Stein, Charles},
  journal={The Annals of Mathematical Statistics},
  volume={16},
  number={3},
  pages={243--258},
  year={1945},
  publisher={JSTOR}
}

@article{wittes1999internal,
  title={Internal pilot studies I: type I error rate of the naive t-test},
  author={Wittes, Janet and Schabenberger, Oliver and Zucker, David and Brittain, Erica and Proschan, Michael},
  journal={Statistics in Medicine},
  volume={18},
  number={24},
  pages={3481--3491},
  year={1999},
  publisher={Wiley Online Library}
}

@article{zucker1999internal,
  title={Internal pilot studies II: comparison of various procedures},
  author={Zucker, David M and Wittes, Janet T and Schabenberger, Oliver and Brittain, Erica},
  journal={Statistics in Medicine},
  volume={18},
  number={24},
  pages={3493--3509},
  year={1999},
  publisher={Wiley Online Library}
}

\clearpage

\begin{table}[]
\caption[]
    {\textit{Summary of variance estimators}}
    \begin{tabular}{lll}
    Estimate & Description & Definition \\
    \hline
$\hat{\sigma}^2_{OS}$&One-sample variance& \eqref{OS-variance} \\
$\hat{\sigma}^2_{ADJ}$&Adjusted one-sample variance&$\hat{\sigma}^2_{OS}-\delta^2 n_{1,int}n_{0,int}/(n_{int}(n_{int}-1))$\\
$\hat{\sigma}^2_{U,1-\gamma}$&Infeasible upper confidence limit&$\hat{\sigma}^2_{OS}(n_{int}-1)/d_{\lambda,1-\gamma}$\\
$\bar{\sigma}^2_{U,1-\gamma}$&Conservative upper confidence limit&$\hat{\sigma}^2_{OS}(n_{int}-1)/d_{1-\gamma}$\\
\hline
\end{tabular}
\end{table}

\begin{table}[]
\caption[]
    {\textit{Confidence level $1-\gamma$ derived with \eqref{lowerbound} under blinded review with balanced allocation ($n_{1,int}=n_{0,int}$), for one-sided significance level $\alpha=0.025$ and the target power $1-\beta=0.80,0.90$}}
    \begin{tabular}{c|cccccccccccc}
    \hline
\multicolumn{1}{c|}{$n_{z,int}$}&2&3&4&5&6&7&8&9&10&20&30&40\\ \hline
\multirow{1}{*}{$1-\beta=0.80$}&0.65&0.62&0.61&0.60&0.59&0.59&0.58&0.58&0.57&0.55&0.54&0.54\\
\multirow{1}{*}{$1-\beta=0.90$}&0.76&0.72&0.69&0.67&0.66&0.65&0.64&0.63&0.62&0.59&0.57&0.57\\

   \hline
    \end{tabular}
\end{table}

\clearpage

\begin{sidewaystable}
\caption
    {Power (\%) with the re-calculated sample size based on the internal pilot study with $n_{z,int}$ subjects for the cases of $N=32,63$ and $174$ ($\sigma^2=2.038, 4.013$ and $11.08$); {\it One-sample}: \eqref{nfin_total} with \eqref{OS-variance}, {\it IF}: \eqref{nfin_zucker}, {\it Proposed}: \eqref{nfin_cons}, {\it Theoretical}: \eqref{nfin_accu}
    }
\centering
\begin{minipage}[t]{0.48\linewidth}
\centering
\begin{tabular}{cccccc}
\hline
 & &\multicolumn{4}{c}{Power}\\
       $N$&$n_{z,int}$ & {\it One-sample} & {\it IF} &{\it Proposed}&{\it Theoretical}\\ 
\hline
32&2&66.28&91.41&80.85&77.48 \\ 
&3&71.15&85.92&81.01&77.28 \\ 
&4&73.65&84.09&81.42&77.53 \\ 
&5&75.17&83.28&81.53&77.53 \\ 
&6&76.20&82.84&81.51&77.41 \\ 
&7&76.96&82.57&81.42&77.25 \\ 
&8&77.52&82.38&81.61&77.40 \\ 
&9&77.97&82.26&81.46&77.20 \\ 
&10&78.32&82.17&81.29&77.30 \\ 
&20&80.02&82.88&81.66&77.32 \\ 
&30&81.72&82.66&82.54&79.60 \\ 
&&&&& \\

63&2&65.61&91.16&80.38&78.66 \\ 
&3&70.45&85.46&80.45&78.53 \\ 
&4&72.93&83.55&80.82&78.82 \\ 
&5&74.44&82.68&80.90&78.83 \\ 
&6&75.46&82.20&80.85&78.73 \\ 
&7&76.20&81.90&80.74&78.58 \\ 
&8&76.76&81.70&80.91&78.73 \\ 
&9&77.20&81.56&80.74&78.53 \\ 
&10&77.55&81.46&80.57&78.64 \\ 
&20&79.17&81.08&80.86&78.56 \\ 
&30&79.71&80.98&80.83&78.50 \\ 
&40&79.99&80.94&80.92&78.58 \\ 
&50&80.25&80.98&80.89&78.66 \\ 
&60&81.19&81.65&81.62&80.06 \\ 

\hline
\end{tabular}
\end{minipage}%
\hfill
\begin{minipage}[t]{0.48\linewidth}
\centering
\begin{tabular}{cccccc}
\hline
 & &\multicolumn{4}{c}{Power} \\
       $N$&$n_{z,int}$ & {\it One-sample} & {\it IF} &{\it Proposed}&{\it Theoretical}\\
\hline
174&2&65.31&91.04&80.15&79.53 \\ 
&3&70.09&85.21&80.15&79.45 \\ 
&4&72.53&83.23&80.47&79.74 \\ 
&5&74.02&82.32&80.52&79.76 \\ 
&6&75.02&81.81&80.45&79.67 \\ 
&7&75.75&81.50&80.32&79.53 \\ 
&8&76.29&81.28&80.48&79.68 \\ 
&9&76.73&81.13&80.30&79.49 \\ 
&10&77.08&81.02&80.12&79.60 \\ 
&20&78.66&80.60&80.37&79.53 \\ 
&30&79.12&80.46&80.33&79.47 \\ 
&40&79.47&80.43&80.40&79.55 \\ 
&50&79.63&80.40&80.30&79.44 \\ 
&60&79.74&80.38&80.34&79.47 \\ 
&75&79.85&80.36&80.36&79.50 \\ 
&100&79.96&80.34&80.38&79.51 \\ 
&125&80.03&80.33&80.30&79.43 \\ 
&150&80.11&80.35&80.35&79.50 \\ 
\hline
&&\\
&&\\
&&\\
&&\\
&&\\
&&\\
&&\\
&&\\
\end{tabular}
\end{minipage}
\end{sidewaystable}

\clearpage

\begin{sidewaystable}
\caption{Type I error rate (\%) with the re-calculated sample size based on the internal pilot study with $n_{z,int}$ subjects for the cases of $N=32,63$ and $174$ ($\sigma^2=2.038, 4.013$ and $11.08$); {\it One-sample}: \eqref{nfin_total} with \eqref{OS-variance}, {\it IF}: \eqref{nfin_zucker}, {\it Proposed}: \eqref{nfin_cons}, {\it Theoretical}: \eqref{nfin_accu}
    }
\centering
\begin{minipage}[t]{0.48\linewidth}
\centering
\begin{tabular}{cccccc}
\hline
 & &\multicolumn{4}{c}{Type I error rate}\\
       $N$&$n_{z,int}$ & {\it One-sample} & {\it IF} &{\it Proposed}&{\it Theoretical}\\ 
\hline
32&2&2.416&2.453&2.433&2.429 \\ 
&3&2.457&2.462&2.459&2.458 \\ 
&4&2.471&2.473&2.472&2.471 \\ 
&5&2.479&2.479&2.481&2.481 \\ 
&6&2.485&2.485&2.487&2.484 \\ 
&7&2.488&2.489&2.486&2.488 \\ 
&8&2.494&2.492&2.491&2.493 \\ 
&9&2.495&2.496&2.495&2.496 \\ 
&10&2.495&2.492&2.495&2.494 \\ 
&20&2.504&2.503&2.504&2.505 \\ 
&30&2.505&2.505&2.505&2.505 \\ 
&&&&& \\

63&2&2.435&2.470&2.453&2.449 \\ 
&3&2.466&2.473&2.471&2.470 \\ 
&4&2.476&2.480&2.478&2.477 \\ 
&5&2.483&2.485&2.483&2.483 \\ 
&6&2.488&2.489&2.488&2.488 \\ 
&7&2.488&2.488&2.488&2.488 \\ 
&8&2.492&2.495&2.492&2.493 \\ 
&9&2.493&2.495&2.496&2.500 \\ 
&10&2.492&2.491&2.491&2.492 \\ 
&20&2.500&2.498&2.498&2.499 \\ 
&30&2.501&2.500&2.501&2.501 \\ 
&40&2.501&2.501&2.501&2.501 \\ 
&50&2.503&2.502&2.503&2.503 \\ 
&60&2.503&2.503&2.503&2.504 \\ 

\hline
\end{tabular}
\end{minipage}%
\hfill
\begin{minipage}[t]{0.48\linewidth}
\centering
\begin{tabular}{cccccc}
\hline
 & &\multicolumn{4}{c}{Type I error rate} \\
       $N$&$n_{z,int}$ & {\it One-sample} & {\it IF} &{\it Proposed}&{\it Theoretical}\\
\hline
174&2&2.463&2.483&2.474&2.475 \\ 
&3&2.481&2.487&2.485&2.485 \\ 
&4&2.486&2.489&2.488&2.487 \\ 
&5&2.489&2.491&2.491&2.490 \\ 
&6&2.494&2.494&2.495&2.495 \\ 
&7&2.493&2.494&2.492&2.493 \\ 
&8&2.498&2.497&2.497&2.498 \\ 
&9&2.498&2.498&2.498&2.499 \\ 
&10&2.494&2.495&2.494&2.493 \\ 
&20&2.497&2.497&2.497&2.498 \\ 
&30&2.498&2.498&2.498&2.498 \\ 
&40&2.500&2.499&2.499&2.499 \\ 
&50&2.501&2.500&2.500&2.500 \\ 
&60&2.500&2.500&2.500&2.500 \\ 
&75&2.502&2.502&2.502&2.503 \\ 
&100&2.500&2.501&2.500&2.501 \\ 
&125&2.501&2.501&2.501&2.502 \\ 
&150&2.502&2.502&2.503&2.502 \\ 
\hline
&&\\
&&\\
&&\\
&&\\
&&\\
&&\\
&&\\
&&\\
\end{tabular}
\end{minipage}
\end{sidewaystable}

\begin{sidewaystable}
\caption
    {Mean (standard deviation) of the re-calculated sample size based on the internal pilot study with $n_{z,int}$ subjects for the cases of $N=32,63$ and $174$ ($\sigma^2=2.038, 4.013$ and $11.08$); {\it One-sample}: \eqref{nfin_total} with \eqref{OS-variance}, {\it IF}: \eqref{nfin_zucker}, {\it Proposed}: \eqref{nfin_cons}, {\it Theoretical}: \eqref{nfin_accu}
    }
\centering
\small
\begin{minipage}[t]{0.48\linewidth}
\centering

\begin{tabular}{cccccc}
\hline
 & &\multicolumn{4}{c}{Mean (standard deviation) of $\hat{n}_{z,fin}$}\\
       $N$&$n_{z,int}$ & {\it One-sample} & {\it IF} &{\it Proposed}&{\it Theoretical}\\ 
\hline
32&2&37.25 (30.09)&136.5 (110.3)&68.06 (55.01)&58.06 (46.92)\\
&3&36.72 (23.02)&64.64 (40.54)&52.11 (32.69)&45.23 (28.37)\\
&4&36.49 (19.35)&52.25 (27.72)&47.21 (25.04)&41.27 (21.89)\\
&5&36.36 (17.02)&47.29 (22.13)&44.48 (20.82)&39.05 (18.28)\\
&6&36.28 (15.36)&44.63 (18.90)&42.71 (18.08)&37.59 (15.92)\\
&7&36.23 (14.11)&42.98 (16.74)&41.43 (16.14)&36.54 (14.23)\\
&8&36.19 (13.12)&41.85 (15.18)&40.85 (14.82)&36.07 (13.08)\\
&9&36.16 (12.32)&51.03 (13.98)&40.03 (13.64)&35.39 (12.06)\\
&10&36.13 (11.64)&40.42 (13.03)&39.36 (12.69)&35.12 (11.32)\\
&20&36.04 (8.067)&37.96 (8.517)&37.73 (8.463)&33.53 (7.458)\\
&30&36.54 (5.815)&37.63 (6.212)&37.50 (6.167)&34.10 (4.699)\\ 
&&&&& \\

63&2&68.25 (55.55)&250.1 (203.6)&124.7 (101.5)&114.9 (93.54)\\
&3&67.71 (42.73)&119.2 (75.23)&96.12 (60.65)&89.32 (56.36)\\
&4&67.49 (35.99)&96.64 (51.54)&87.31 (46.57)&81.44 (43.43)\\
&5&67.36 (31.69)&87.61 (41.21)&82.41 (38.77)&77.02 (36.23)\\
&6&67.29 (28.63)&82.77 (35.22)&79.19 (33.70)&74.11 (31.54)\\
&7&67.23 (26.32)&79.76 (31.22)&76.88 (30.10)&72.01 (29.19)\\
&8&67.19 (24.49)&77.70 (28.32)&75.84 (27.64)&71.09 (25.91)\\
&9&67.16 (22.99)&76.22 (26.10)&74.36 (25.46)&69.73 (23.87)\\
&10&67.13 (21.74)&75.09 (24.32)&73.13 (23.68)&69.20 (22.41)\\
&20&67.03 (15.15)&70.61 (15.96)&70.18 (15.87)&65.96 (14.91)\\
&30&66.99 (12.31)&69.31 (12.74)&69.03 (12.69)&64.91 (11.93)\\
&40&66.98 (10.63)&68.69 (10.90)&68.64 (10.90)&64.57 (10.24)\\
&50&67.04 (9.335)&68.38 (9.569)&68.21 (9.540)&64.27 (8.796)\\
&60&67.90 (7.363)&68.84 (7.707)&68.77 (7.683)&65.60 (6.337)\\

\hline
\end{tabular}
\end{minipage}%
\hfill
\begin{minipage}[t]{0.48\linewidth}
\centering
\begin{tabular}{cccccc}
\hline
 & &\multicolumn{4}{c}{Mean (standard deviation) of $\hat{n}_{z,fin}$} \\
       $N$&$n_{z,int}$ & {\it One-sample} & {\it IF} &{\it Proposed}&{\it Theoretical}\\
\hline
174&2&179.3 (146.3)&657.0 (536.2)&327.6 (267.4)&317.9 (259.5)\\
&3&178.7 (113.0)&314.7 (199.0)&253.7 (160.4)&247.0 (156.2)\\
&4&178.5 (95.37)&255.6 (136.6)&230.9 (123.4)&225.1 (120.3)\\
&5&178.4 (84.06)&232.0 (109.3)&218.2 (102.8)&212.8 (100.3)\\
&6&178.3 (76.01)&219.3 (93.49)&209.8 (89.46)&204.8 (87.30)\\
&7&178.2 (69.89)&211.4 (82.92)&203.8 (79.93)&199.0 (78.03)\\
&8&178.2 (65.06)&206.1 (75.24)&201.1 (73.43)&196.4 (71.70)\\
&9&179.2 (61.10)&202.2 (69.34)&197.3 (67.65)&192.7 (66.07)\\
&10&179.1 (57.79)&199.3 (64.63)&194.1 (62.95)&191.2 (62.01)\\
&20&178.0 (40.31)&187.6 (42.47)&186.4 (42.21)&182.2 (41.25)\\
&30&178.0 (32.76)&184.1 (33.90)&183.4 (33.76)&179.3 (33.00)\\
&40&178.0 (28.31)&182.5 (29.04)&182.4 (29.02)&178.3 (28.37)\\
&50&178.0 (25.29)&181.6 (25.80)&181.1 (25.74)&177.1 (25.16)\\
&60&178.0 (23.07)&180.9 (23.45)&180.7 (23.43)&176.7 (22.90)\\
&75&178.0 (20.61)&180.3 (20.89)&180.3 (20.89)&176.3 (20.42)\\
&100&177.9 (17.84)&179.7 (18.01)&179.9 (18.03)&175.9 (17.63)\\
&125&177.9 (15.94)&179.3 (16.07)&179.2 (16.06)&175.3 (15.70)\\
&150&178.1 (14.31)&179.2 (14.45)&179.2 (14.44)&175.3 (13.96)\\
\hline
&&\\
&&\\
&&\\
&&\\
&&\\
&&\\
&&\\
&&\\
\end{tabular}
\end{minipage}
\end{sidewaystable}

\clearpage

\begin{figure}[htbp]
  \centering
  
  \begin{subfigure}{0.55\textwidth}
   \caption{$N=32$}
    \centering
    \includegraphics[width=\linewidth]{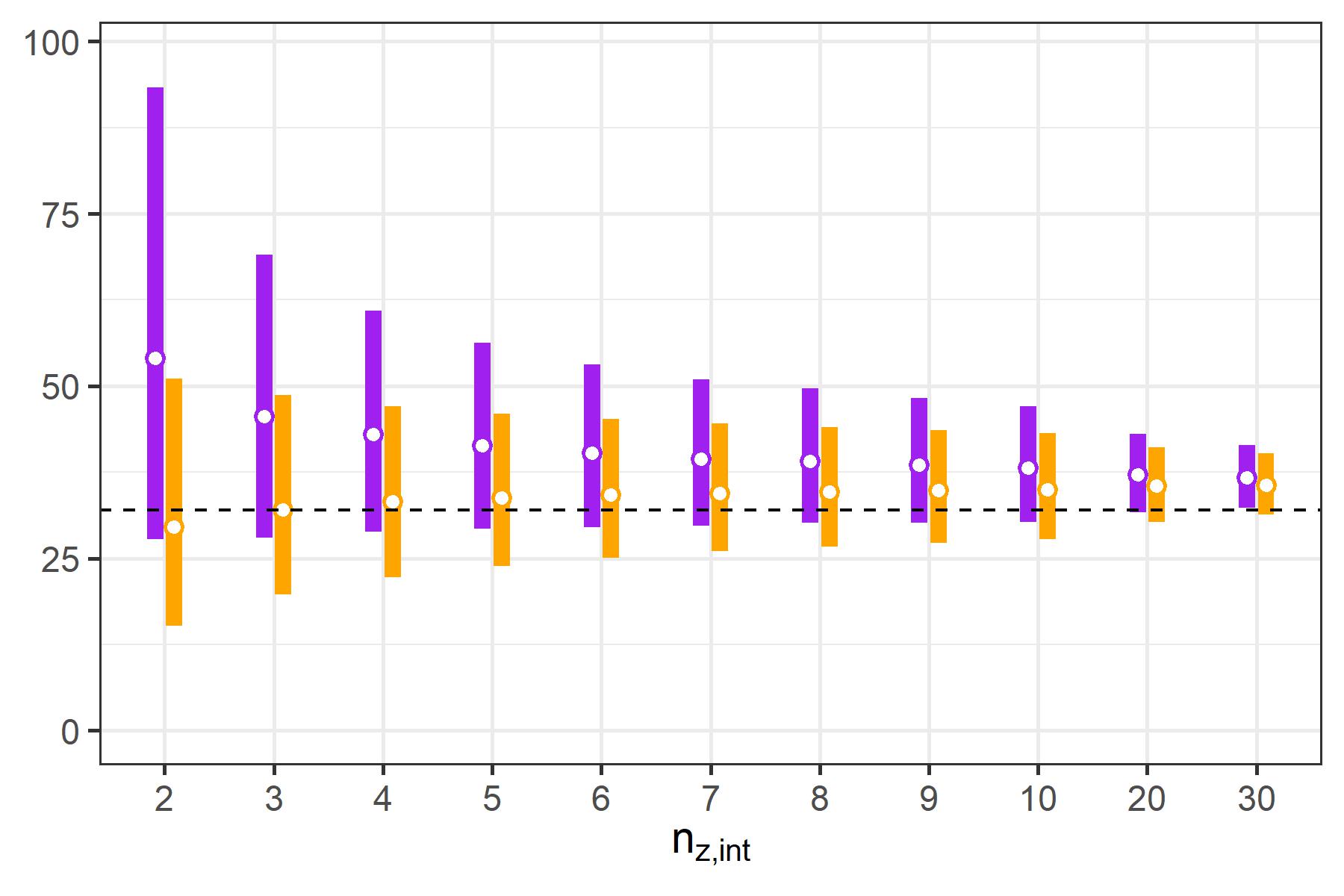}
   
  \end{subfigure}
  
  \vspace{0.8em} 
  
  \begin{subfigure}{0.55\textwidth}   
  \caption{$N=63$}
    \centering
    \includegraphics[width=\linewidth]{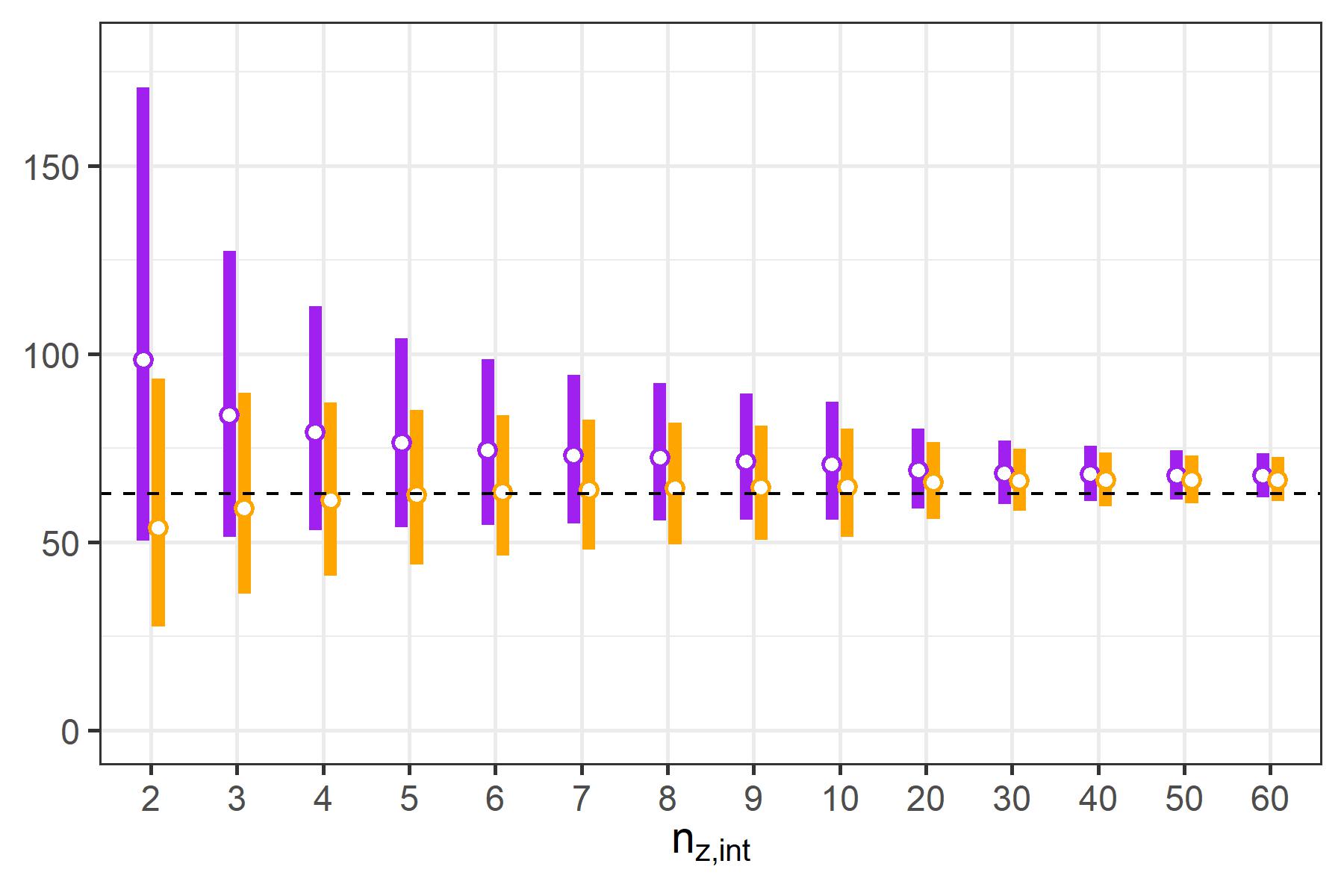}

  \end{subfigure}
  
  \vspace{0.8em} 
  
  \begin{subfigure}{0.55\textwidth}
      \caption{$N=174$}
    \centering
    \includegraphics[width=\linewidth]{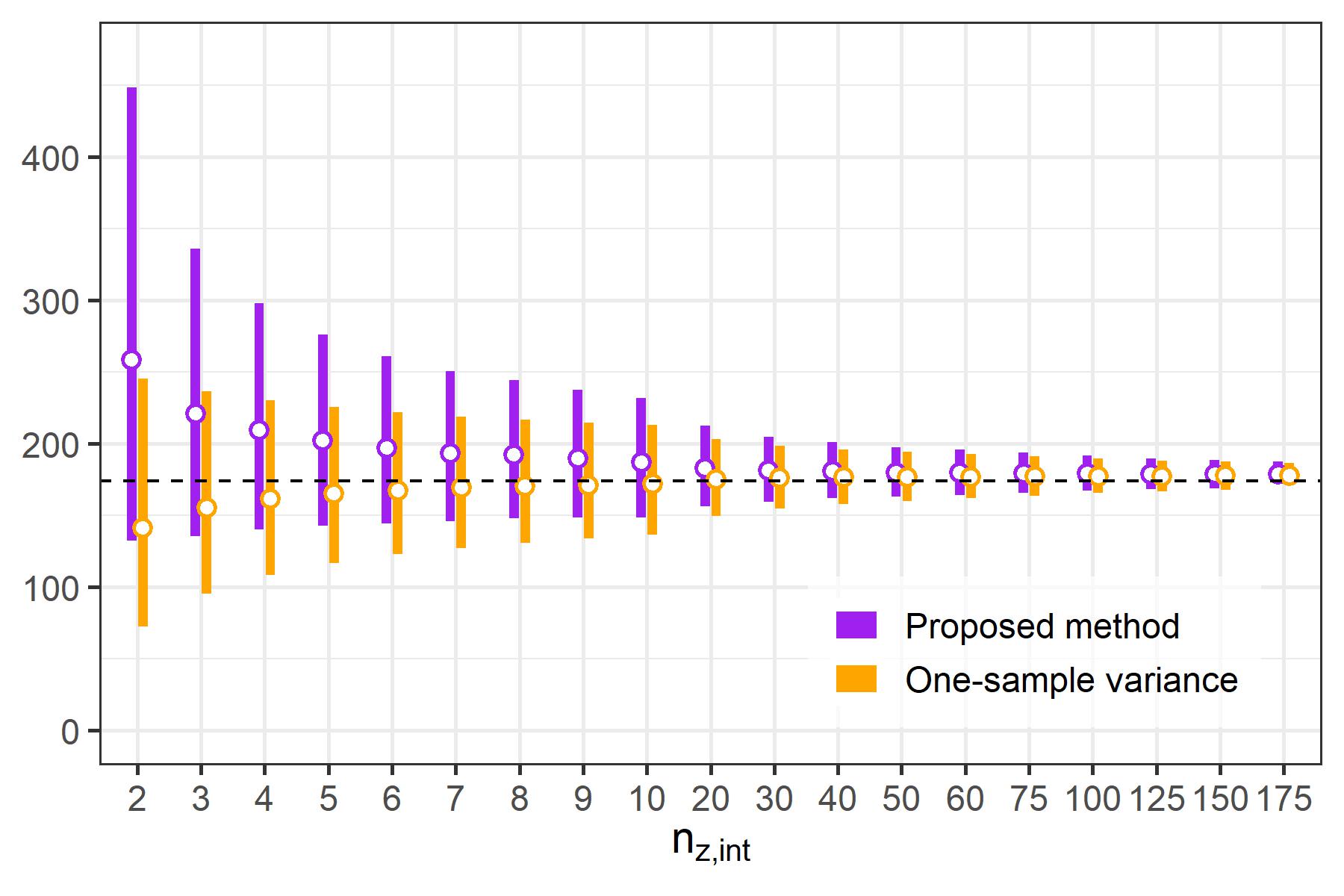}

  \end{subfigure}
  
  \caption{
  Distribution (median with the $1^{st}$ and $3^{rd}$ quartiles) of the re-calculated $\hat{n}_{z,fin}$ with the {\it Proposed} and the {\it One-sample} methods under balanced allocation ($n_{1,int}=n_{0,int},\pi=0.5$) for the cases of (a) $N=32$, (b) $N=63$, and (c) $N=174$.
  }
  \label{fig:nfin_compare}
\end{figure}

\end{document}